\documentclass[12pt]{iopart}
\usepackage{amssymb,latexsym,mathrsfs}
\usepackage{amsfonts}
\usepackage{graphicx}
\usepackage{color}    
\usepackage{textcomp}

\newcommand{\be}{\begin{equation}}
\newcommand{\ee}{\end{equation}}
\newcommand{\bearr}{\begin{array}}
\newcommand{\enarr}{\end{array}}

\def\bea{\begin{eqnarray}}
\def\eea{\end{eqnarray}}
\def\ba{\begin{array}}
\def\ea{\end{array}}

\begin{document}

\title{Particle hopping on a ladder: exact solution using multibalance}
\author{Indranil Mukherjee}
\address{Department of Physical Sciences, Indian Institute of Science Education and Research Kolkata, Mohanpur - 741246, India.}
\ead{im20rs148@iiserkol.ac.in}

\begin{abstract}
We study particle hopping on a two-leg ladder where a  particle   can jump to  their immediate  neighbours, one at  a time, with rates that depend on the occupation  of the departure site and a neighbouring site  on the  other leg. For specific choices of rates, the model  can be solved using pairwise balance  known earlier. For the other  regimes,  we  introduce  a new balance condition called multibalance  which helps us  in obtaining  the exact steady state.  The direction  of the total  current in these models does not necessarily decide the direction of the currents  in individual legs; we  find the regions in the parameter space  where the currents in individual 
legs alter their direction.  In some parameter regime, the total current  exhibits the re-entrance phenomena, 
in   the sense  that  the total current flips  its direction  with increase  of certain parameter and  flips it again  when the parameter is  increased further. It  turns out that the  multibalance  condition we introduce here is  very useful and it can be  applied  generically  to several other models. We discuss  some of these  models in short.
\end{abstract}
\maketitle

\section{Introduction}

Nonequilibrium steady states (NESS) \cite{DDS,book} differ from their equilibrium counterparts
which obey detailed balance \cite{book_db_1,book_db_2}. Detailed balance ensures that there is no net flow of probability current among any pair of configurations leading to the well known
Gibbs-Boltzmann measure in its steady state. Such a generic measure is absent in nonequilibrium  and in general, finding an exact NESS measure for any nonequilibrium dynamics is usually difficult.
It has been realized that exact solutions of steady state measures for certain non-equilibrium systems and analytical calculation of observables bring much insight to
the understanding of the corresponding systems. In context of the exactly solvable
interacting non-equilibrium systems, there exist  a few successful models. The zero range process (ZRP) \cite{Spitzer, Evans_Braz, Evans_Hanney} and certain   lattice gas models  in one dimension are  perhaps the simplest of
them, which exhibit nontrivial static and dynamic properties in the steady state. These models have  found applications  in describing phase separation criteria in driven lattice gases \cite{Kafri_Mukamel}, network re-wiring \cite{Angel_Evans, PKM_JALAN}, statics and dynamics of extended objects \cite{Gupta_Barma_PKM, Daga_PKM}. etc. 

The steady state of the well studied ZRP  can be  obtained  using  a   pairwise balance (PWB)  condition \cite{PWB} where  for every transition  $C\to C'$   one needs to find a  unique configuration $C''$ such that the out-flux  from  $C$ to $C'$ is
balanced by the in-flux from $C''$ to $C$.
The steady state of asymmetric simple exclusion process  (ASEP) also follows   the PWB condition.   These models  with open boundaries 
could be solved  exactly \cite{derrida__tasep_mpa} using a matrix product ansatz (MPA),  where the steady state weight of any configuration is represented by a product of non-commuting matrices.  This matrix product ansatz  has been successfully   implemented  in several other models.  Some examples  include ASEP with open boundaries \cite{derrida__tasep_mpa, evans_mpa},  multiple species
of particles \cite{EVANS_MR} and  models where particles have some  internal degrees of freedom \cite{UBASU_PKM_TASEP}, non-conserved systems with deposition, evaporation, coagulation-decoagulation like dynamics \cite{Hinrichsen}.
Another class of nonequilibrium models, finite range processes (FRP) have been studied recently \cite{FRP}. It is shown that, for certain specific conditions on hop rates, the FRP has a cluster-factorized steady state (CFSS). The steady states of these models can  be achieved by both pairwise balance and $h$-balance conditions \cite{FRP, asymm_FRP} and they  exhibit a finite dimensional transfer-matrix representation of the steady state. 

In this article we introduce  another balance condition, namely multibalance (MB), to non-equilibrium  steady states  (NESS): for  every configuration $C,$ the sum of outgoing fluxes to  one or more  configurations are  balanced here  by   the sum of multiple incoming fluxes. A recent article \cite{barma} has independently discussed this balance condition to exactly determine the steady state of a model namely light-heavy (LH) model. This balance condition is referred to as bunchwise balance.  We applied the  multibalance condition to the  model of particle hopping on a two-leg ladder, where the  hop rates  depend on the  occupation number of the departure  site  and  corresponding occupation number in the other leg.  The steady state currents in this  model  exhibit  several interesting features.  When  some of the parameters of the model
increased, the total current  flipped its direction and  further increase of  the same parameter resulted in another flip  of direction. Along with this re-entrance phenomena, we  also find that the direction of the   total  current does not necessarily   dictate the direction of the  currents in individual legs; 
we explicitly  obtain  the line of separation  where currents in individual legs alter their direction. 

The new balance  condition turns out to be a very useful  method to  obtain exact steady state of  some  nonequilibrium models.  It can  be  implemented to obtain factorized or pair-factorized steady states  (PFSS), which is described in section \ref{sec:FSS} and \ref{sec:PFSS}.  We generalize the  hopping rates  of the two-leg ladder   to  obtain   a PFSS  when  the rates   satisfy the multibalance conditions.  More generic models, like  finite range processes   which give rise to cluster-factorized steady states (CFSS)  and   systems  with more than one species of particles  are discussed briefly  in   section \ref{sec:CFSS}; exact steady states   obtained   for these models clearly  emphasize the  utility and strength of this new balance condition.  
  
\section{Multibalance (MB)}

We define a generalized balance condition  in nonequilibrium systems such that a bunch of fluxes coming to the configuration $C$ from a  set of configurations  $\{ C''_{1}, \cdots, C''_{N_{C}}\}$ are balanced by the sum of out-fluxes from $C$  to a  set of configurations $\{C'_{1}, \cdots, C'_{M_{C}}\}$ in the configuration space. 
\begin{figure}[h]
\vspace*{.5 cm}
\centering
\includegraphics[height=4.0 cm]{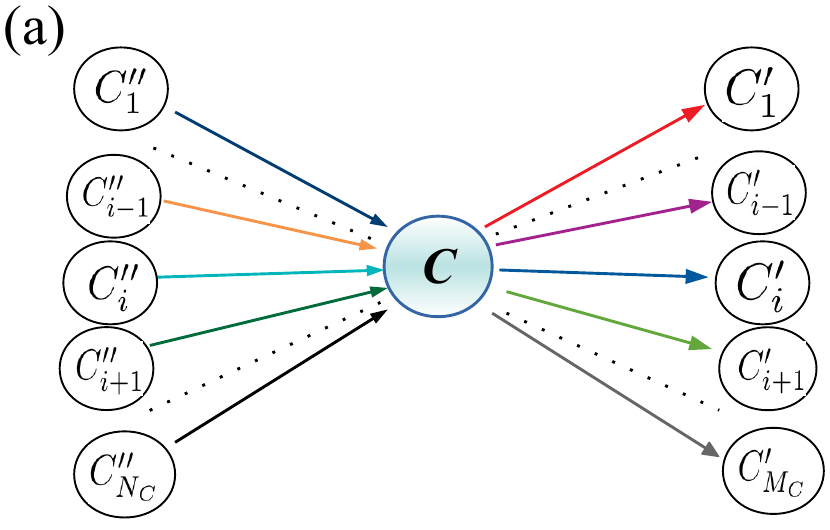} \hspace{.3 cm}\includegraphics[height=4.0 cm]{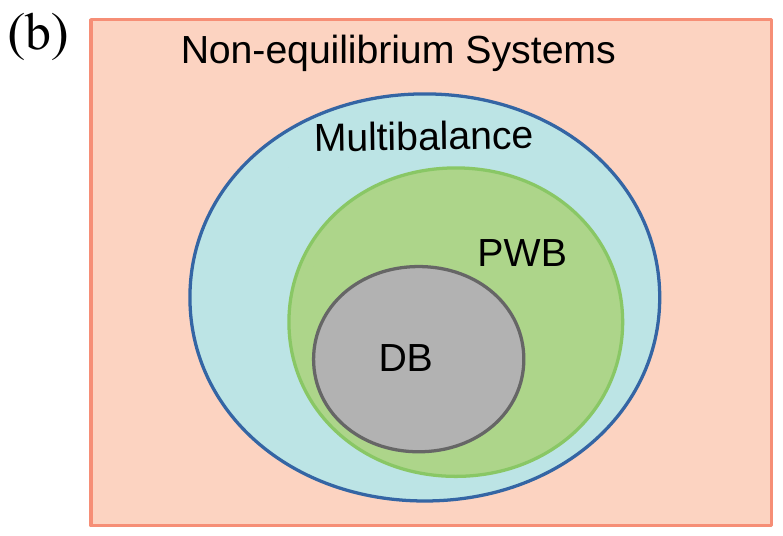}
\caption{(a) Multibalance: fluxes are represented by arrows. Incoming fluxes to the configuration $C$ from a set of configurations  $\{ C''_{1}, \cdots,  C''_{N_{C}}\}$ are balanced with the outward fluxes from $C$ to the set of configurations $\{C'_{1}, ...,  C'_{M_{C}}\}$, (b) In nonequilibrium systems, multibalance is a generalised balance condition to obtain the steady state. When $N_{C} = M_{C} = 1$, PWB is a subset of multibalance for  
$C''_{1} \neq C_{1}'$ and DB is a subset of PWB for $ C''_{1} = C_{1}'$, corresponds to the equilibrium case.}
 \label{fig:multibalance}
\end{figure}
Here $N_{C}$ is the total number of incoming fluxes for the set of configurations $\{ C''_{1}, \cdots,  C''_{N_{C}}\}$ and $M_{C}$ is the total number of outgoing fluxes for the set of configurations $\{C'_{1}, ...,  C'_{M_{C}}\}$ as described in Fig. \ref{fig:multibalance}(a). At steady state,  for any system,   the  fluxes must balance:  $\sum_{C'} P(C) W(C \rightarrow C') = \sum_{C''} P( C'') W( C'' \rightarrow C)$. We have denoted $P(C)$ be the probability of the configuration $C$ and it can move to the other configuration $C'$ with a dynamical rate $W(C\rightarrow C')$.    For systems    that satisfy a multibalance, these  steady state configurations  break into  many  conditions of the form,  
\begin{eqnarray} 
\sum_{i=1}^{M_{C}} P(C) W(C \rightarrow C'_{i})
 = \sum_{i=1}^{N_{C}} P( C''_{i}) W( C''_{i} \rightarrow C).\label{eq:master_MB}
\end{eqnarray}
Eq. (\ref{eq:master_MB}) describes that for every configuration $C$, the incoming fluxes from a group of configurations $\{ C''_{1}, \cdots,  C''_{N_{C}}\}$, are balanced by outgoing fluxes to another uniquely identified group of configurations $\{C'_{1}, ...,  C'_{M_{C}}\}$. As a special case of multibalance condition, for $N_{C} = M_{C} = 1$, if   
$C''_{1} \neq C_{1}'$, Eq. (\ref{eq:master_MB}) reduces to Pairwise balance balance condition (PWB) and for the simplest case when $C''_{1} = C_{1}'$, it becomes the well known Detailed balance condition (DB) corresponds to the  equilibrium case as shown in  Fig. \ref{fig:multibalance}(b).

\subsection{Zero range process (ZRP) in two dimensions  with asymmetric rates \label{sec:2DZRP}}

The zero range process (ZRP) is a model in which  many indistinguishable particles occupy sites on a lattice and these particles hop between neighbouring sites with a rate that depends on the number of particles at the site of departure.  The steady state of ZRP can be solved exactly  in any dimension for periodic boundaries and   in some cases, for  open  boundaries \cite{Evans_Hanney, open_ZRP_Levine, mpa_zrp}.
\begin{figure}[h]
\begin{center}
\vspace*{.5 cm}
\includegraphics[height=5.0 cm]{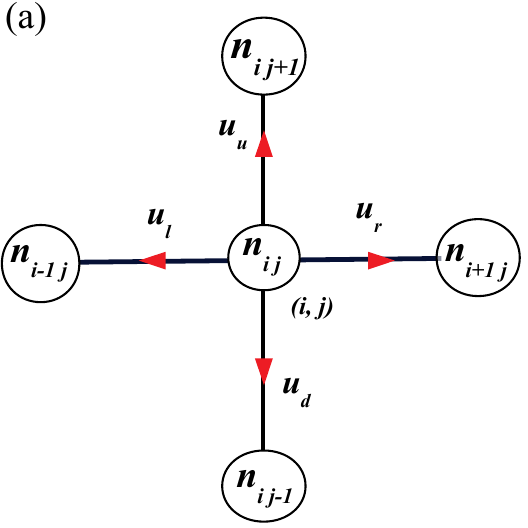} \hspace{.5 cm}\includegraphics[height=5.0 cm]{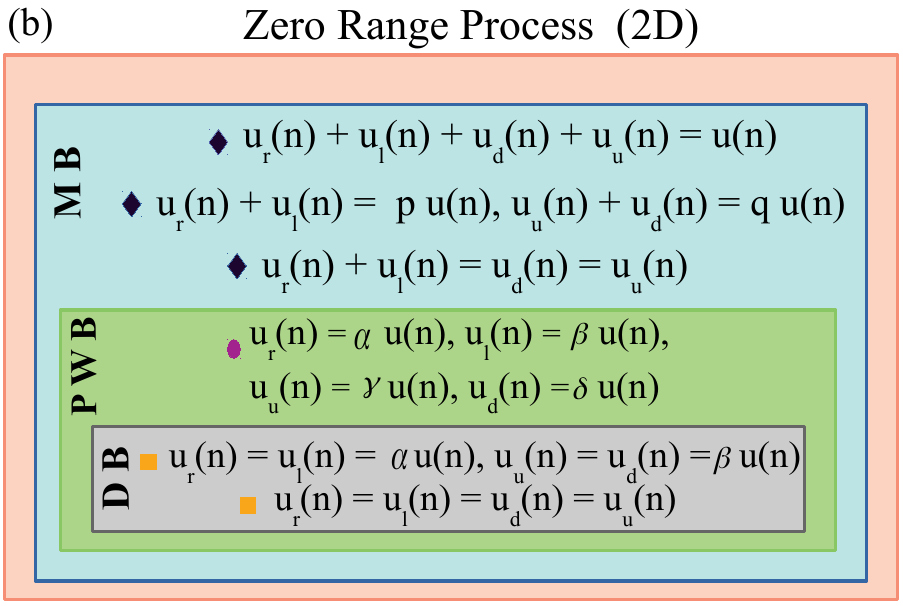}
\end{center}
\caption{(a) ZRP in two dimensions, where a particle from site $(i,j)$ can hop to its right, left, up and down nearest neighbours with rates $u_{r}(n_{i,j})$, $u_{l}(n_{i,j})$, $u_{u}(n_{i,j})$ and $u_{d}(n_{i,j})$ respectively. $n_{i,j}$ is the number of particles at site $(i,j)$, (b) FSS can be obtained for this 2D ZRP model using different balance schemes -  MB, PWB and DB. Examples of corresponding conditions on the hop rates are mentioned there in respective regions and the FSS can be obtained following any of these conditions.  Clearly, MB is the generalized choice to  obtain the FSS  where the hop rates are different in all four directions. }
 \label{fig:2Dzrp}
\end{figure}

We consider a periodic lattice in two dimensions of size ($L \times L$). Each site $(i,j)$  with, $i = 1, 2, \cdots L$, $j = 1, 2, \cdots L$, can be either vacant or it can be occupied by one or more particles $n_{i,j} \leq N$ $\left(N = \sum_{i=1}^{L} \sum_{j=1}^{L} n_{i,j} \right)$. A particle from any of the sites $(i,j)$ can hop to its nearest neighbours (right, left, up and down)  with rates $u_{r}(n_{i,j})$, $u_{l}(n_{i,j})$, $u_{u}(n_{i,j})$ and $u_{d}(n_{i,j})$ respectively as shown in Fig. \ref{fig:2Dzrp}(a).
We assume that the model evolves to a factorized steady state (FSS)
\begin{equation} \label{eq:FSS_P}
 P(\{n_{i,j}\}) \propto \prod_{i=1,j=1}^{L} f(n_{i,j})~ \delta \left (\sum_{i=1,j = 1}^{L} n_{i,j} -N \right ).
\end{equation}
$N$ is the total number of particles and the density of the system $\rho = \frac{N}{L^{2}}$ is conserved by the dynamics. Our task is to verify on what condition we can get FSS as in  Eq. (\ref{eq:FSS_P}) for this model \\
(a) When the rates $u_{r}(n) = u_{l}(n) = \alpha u(n)$ and $u_{u}(n) = u_{d}(n) = \beta u(n)$, where both $\alpha$ and $\beta$ are constants or  $u_{r}(n) = u_{l}(n) = u_{u}(n) = u_{d}(n) = u(n)$, we can obtain the FSS using DB condition. (b) FSS can be obtained using PWB condition when all rates $u_{r}(n)$, $u_{l}(n)$,  $u_{u}(n)$ and $u_{d}(n)$ differ by a multiplicative constant {\it i.e.,} the ratios of the rates are independent of $n$. Steady state weight $f(n) = \prod_{\nu=1}^{n} u(\nu)  ^{-1}$ in both the cases. (c)  It is {\it a priori } not clear, whether a FSS is at all possible for ZRP  
when hop rates in all four directions are different. It is possible to obtain the exact FSS as in Eq. (\ref{eq:FSS_P}) using MB that increases the regime of solvability with any of  the following conditions on hop rates
\begin{eqnarray} \label{eq:FSS_MB}
(i)~~u_{r}(n) + u_{l}(n) + u_{d}(n) + u_{u}(n) = u(n),\label{eq:2dzrp_cond1} \\
(ii)~~u_{r}(n) + u_{l}(n) =p u(n)~{\rm ,}~~ u_{u}(n) + u_{d}(n) =q u(n), \label{eq:2dzrp_cond2}\\
(iii)~~u_{r}(n) + u_{l}(n) = u_{u}(n)=u_{d}(n)= u(n),\label{eq:2dzrp_cond3}
\end{eqnarray}
where $p$ and $q$ in Eq. (\ref{eq:2dzrp_cond2}) are constants. The steady state weight is defined as
$f(n) = \prod_{\nu=1}^{n} u(\nu)^{-1}$. We provide explicit proof, in the Appendix, that, exact FSS can be obtained when the hop rates satisfy Eqs. (\ref{eq:2dzrp_cond1}) - (\ref{eq:2dzrp_cond3}).

\section{Particle hopping on a two-leg ladder \label{sec:FSS}}

\begin{figure}[h]
\vspace*{.5 cm}
\begin{center}
\includegraphics[scale = 0.4]{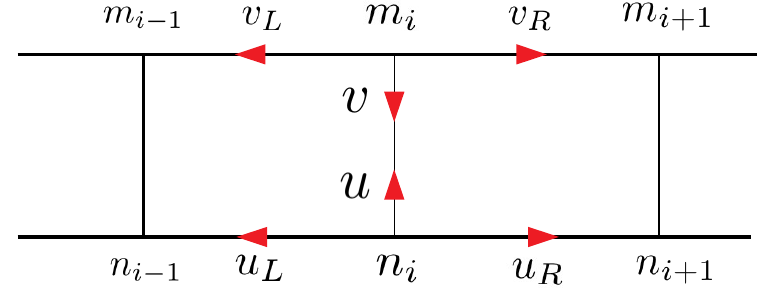}
\end{center}
\caption{Periodic ladder, with two legs, $n_{i}$ and $m_{i}$ are number of particles at site $i$ in the lower and upper leg respectively. A particle from any randomly chosen site $i$ of lower leg, can hop to its right nearest neighbour with rate $u_{R}(n_{i},m_{i})$, left nearest neighbour with rate $u_{L}(n_{i},m_{i})$ and site $i$ of upper leg with rate $u(n_{i},m_{i})$. For upper leg, a particle from site $i$ can hop to its right nearest neighbour with rate $v_{R}(n_{i},m_{i})$, left nearest neighbour with rate $v_{L}(n_{i},m_{i})$ and site $i$ of lower leg with rate $v(n_{i},m_{i})$.}
\label{fig:MB_ladder_pic}
\end{figure}

Let us consider a periodic two-leg ladder (see Fig. \ref{fig:MB_ladder_pic}) with sites at each leg are labeled by $i=1,2,\cdots , L$. For both the legs, each site $i$ can be either vacant or it can be occupied by one or more particles: $n_{i}$ particles in the lower leg and $m_{i}$  in the upper leg. We assume that the hop rates depend on the occupation number of the departure site and  the corresponding site on the 
other leg. From any randomly chosen site $i$ of the lower leg, one particle can hop to its right nearest neighbour with rate $u_{R}(n_{i},m_{i})$, left nearest neighbour with rate $u_{L}(n_{i},m_{i})$ and site $i$ of upper leg with rate $u(n_{i},m_{i})$. Similarly for upper leg, one particle from site $i$, can hop to its right and left nearest neighbours with rate $v_{R}(n_{i},m_{i})$ and $v_{L}(n_{i},m_{i})$ respectively and site $i$ of lower leg with rate $v(n_{i},m_{i})$. 
We demand that the model evolves to a FSS
\begin{eqnarray} 
 P(\{n_{i},m_{i}\}) = \frac{1}{Q_{L,N}} \prod_{i=1}^{L} 
 f(n_{i},m_{i})
~\delta \left(\sum_{i} (n_{i}+m_{i}) - N\right)  \label{eq:MB_ladder_P_FSS}
\end{eqnarray}
with a canonical partition function 
\begin{equation}\label{eq:MB_ladder_QLN_FSS}
Q_{L,N} =\sum_{\{n_{i}\};\{m_{i}\}}\prod_{i=1}^{L} f(n_{i},m_{i})
~\delta \left(\sum_{i} (n_{i}+m_{i}) - N\right).
\end{equation}
The total number of particles 
$N = \sum_{i=1}^L  (m_i +n_i)$ and thus  the density of the system $\rho = \frac{N}{2L}$  is  conserved by the dynamics.
FSS as in Eq. (\ref{eq:MB_ladder_P_FSS}) can be obtained using - (a) DB, when $u_{R}(n,m)=u_{L}(n,m)= \alpha u(n,m)$, $v_{R}(n,m)=v_{L}(n,m)=\beta v(n,m)$, $\alpha$ and $\beta$ are two different constants. (b) PWB, when the rates $u_{R}(n,m) = \alpha u(n,m)$, $u_{L}(n,m)= \beta u(n,m)$, $v_{R}(n,m)=\gamma v(n,m)$ and $v_{L}(n,m)=\delta v(n,m)$ where $\alpha$, $\beta$, $\gamma$ and $\delta$ are four different constants, {\it i.e,} the ratios of the rates are independent of $n$ and $m$. The steady state weight in both the cases is defined by $f(n.m)=\prod_{i=1}^{n} [u(i,m)]^{-1} \prod_{j=1}^{m} [v(0,j)]^{-1}$. Our aim  is to find whether  such a FSS is possible  when rate functions in all hopping directions are different.  To answer this question, we employ the MB condition described in Eq. (\ref{eq:master_MB}).\\
(i) Let us consider a configuration $C \equiv \{ \cdots, n_{i-1},m_{i-1}, n_{i},m_{i},n_{i+1},m_{i+1} \cdots \}$. Fluxes generated by particle hopping from site $i$ of lower leg, to its right and left nearest neighbours of this configuration $C$, can be balanced with  the flux obtained by a particle hopping from site $i$ in the upper leg of another configuration $C' \equiv \{ \cdots, n_{i-1},m_{i-1}, n_{i}-1,m_{i}+1,n_{i+1},m_{i+1} \cdots \}$ to site $i$ of the lower leg. The flux balance scheme in Eq. (\ref{eq:master_MB}) gives the following equation 
\begin{eqnarray}
\fl v(n_{i}-1,m_{i}+1)~P(\cdots, n_{i-1},m_{i-1}, n_{i}-1,m_{i}+1,n_{i+1},m_{i+1} \cdots) \cr \fl =
 [~ u_{R}(n_{i},m_{i}) + u_{L}(n_{i},m_{i})~] ~P(\cdots, n_{i-1},m_{i-1}, n_{i},m_{i},n_{i+1},m_{i+1} \cdots). \label{eq:ladder_FSS_urul}
\end{eqnarray}
(ii) Similarly, fluxes generated by a particle hopping from site $i$ of upper leg to its right and left nearest neighbours of the  configuration $C$, can be balanced with  the flux obtained by a particle hopping from site $i$ in the lower leg of another configuration $C'' \equiv \{ \cdots, n_{i-1},m_{i-1}, n_{i}+1,m_{i}-1,n_{i+1},m_{i+1} \cdots \}$ to site $i$ of the upper leg. Then, the flux balance scheme in Eq. (\ref{eq:master_MB}) gives the following equation
\begin{eqnarray}
\fl u(n_{i}+1,m_{i}-1)~P(\cdots, n_{i-1},m_{i-1}, n_{i}+1,m_{i}-1,n_{i+1},m_{i+1} \cdots) \cr \fl =
 [~ v_{R}(n_{i},m_{i}) + v_{L}(n_{i},m_{i})~] ~P(\cdots, n_{i-1},m_{i-1}, n_{i},m_{i},n_{i+1},m_{i+1} \cdots).\label{eq:ladder_FSS_vrvl}
\end{eqnarray}
One can verify that a factorized form of steady state, as in Eq. (\ref{eq:MB_ladder_P_FSS}), is indeed possible when the hop rates at site $i$ satisfy the following conditions
\begin{equation}
[~ u_{R}(n_{i},m_{i}) + u_{L}(n_{i},m_{i})~] = u(n_{i},m_{i})= \frac{f(n_{i}-1,m_{i})}{f(n_{i},m_{i})},\label{eq:ladder_FSS_con_u}
\end{equation}
\begin{equation}
[~ v_{R}(n_{i},m_{i}) + v_{L}(n_{i},m_{i})~] = v(n_{i},m_{i})= \frac{f(n_{i},m_{i}-1)}{f(n_{i},m_{i})}  \label{eq:ladder_FSS_con_v}
\end{equation}
and the steady state weight is defined as 
\begin{equation} \label{eq:ladder_FSS_weight}
 f(n,m)=\prod_{i=1}^{n} [u(i,m)]^{-1} \prod_{j=1}^{m} [v(0,j)]^{-1}. 
\end{equation}

\subsection{Calculation of current} 

\begin{figure}[h]
\vspace*{.5 cm}
\centering
\includegraphics[height=5.0cm]{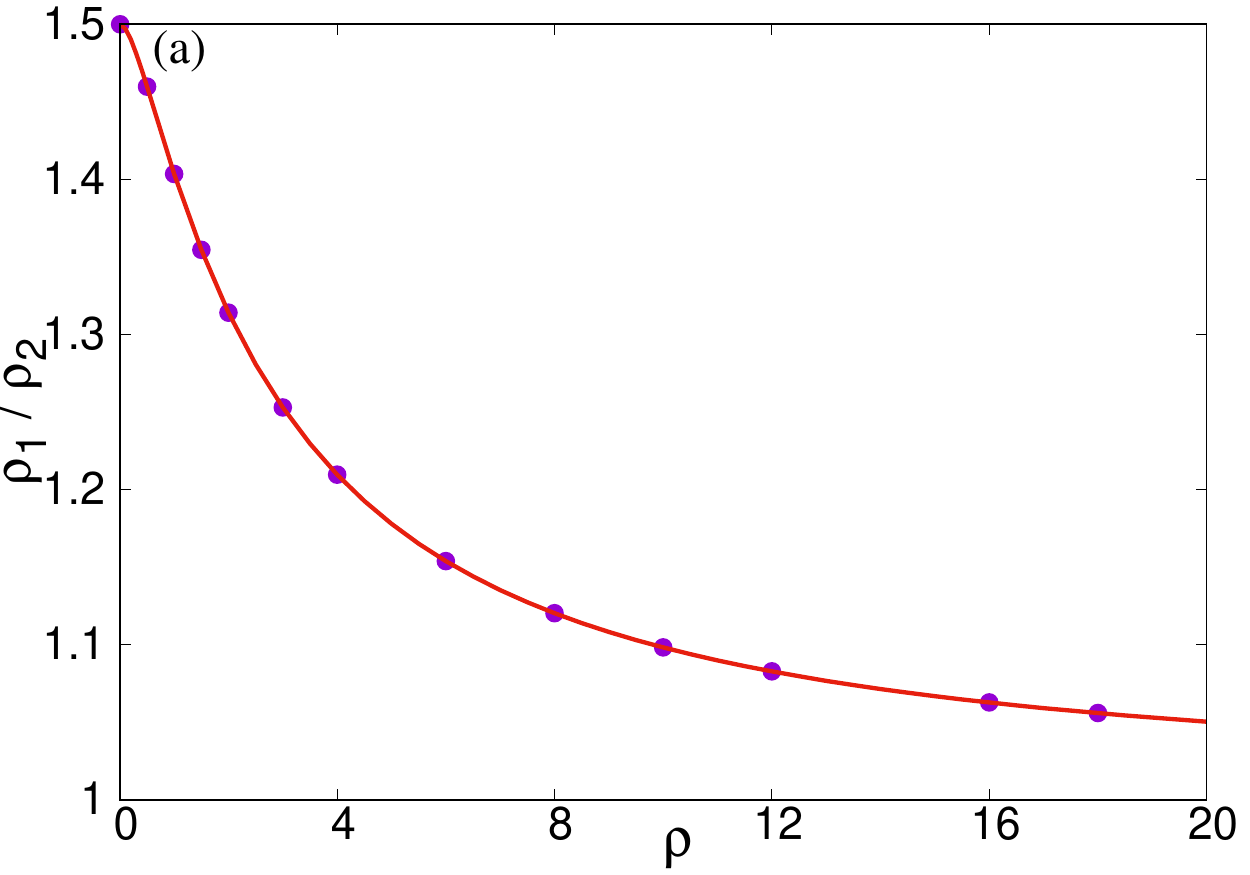}
\hspace{.2 cm}\includegraphics[height=5.0cm]
{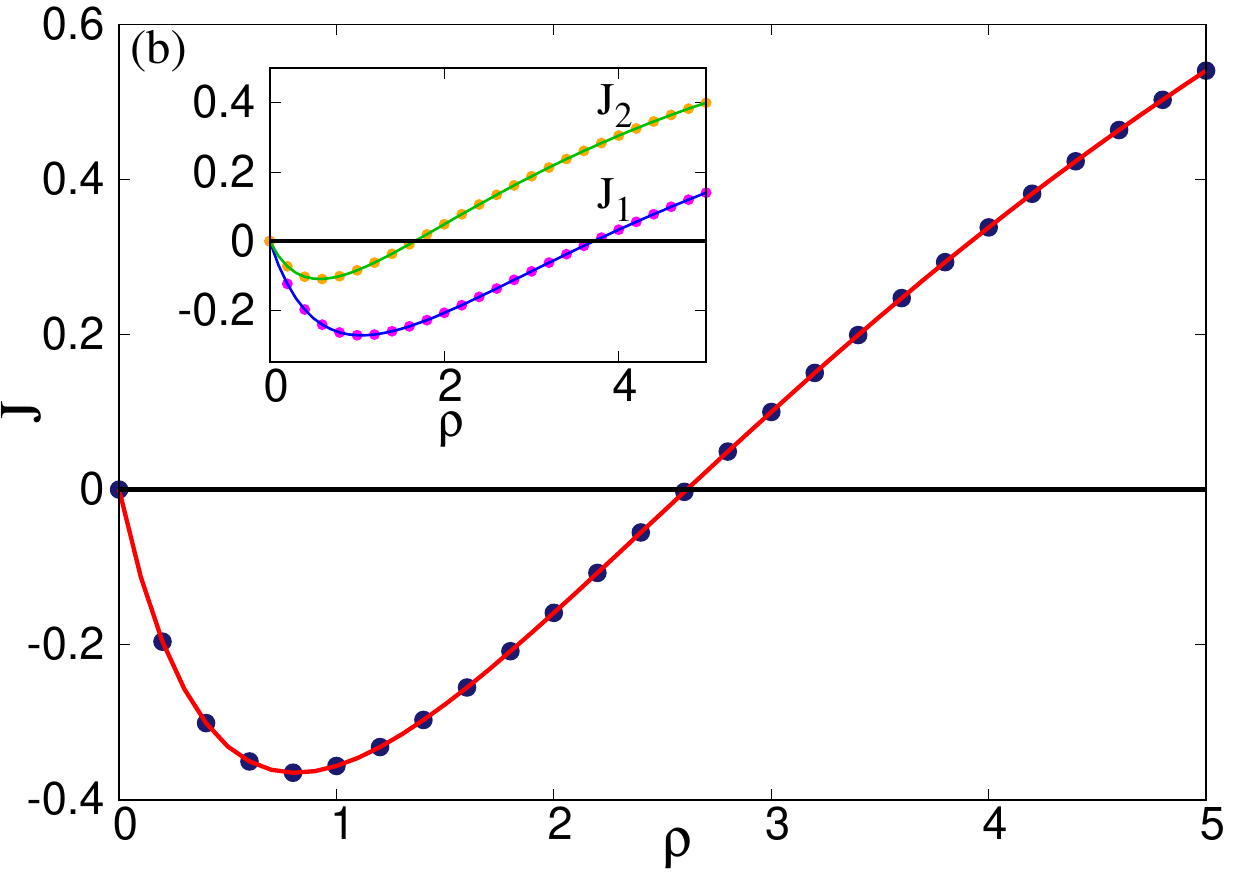}  
 \caption{(a) Relative particle density $\frac{\rho_{1}}{\rho_{2}}$   as a function of density $\rho$, measured from simulation (points) on a system of size $L=100$, is compared with exact expressions of $\rho_{1}$ and $\rho_{2}$ as in Eq. (\ref{eq:fss_rho1_rho2}). (b) The total current $J$ as a function of density $\rho$ for $\alpha=0.6$. The total current reverses its direction at  $\rho=2.611$. The inset shows currents $J_{1,2}$ as a function of $\rho$ for $\alpha=0.6$. Points are from simulations with $L=100$ and averaged over $10^{8}$ trajectories, solid lines are exact according to Eqs. (\ref{eq:ladder_fss_J1}) - (\ref{eq:ladder_fss_J}).}
 \label{fig:ladder_fss_rho1_rho2}
\end{figure}

\begin{figure}[h]
\vspace*{.5 cm}
\centering
 \includegraphics[height=4.95cm,width=5.55cm]{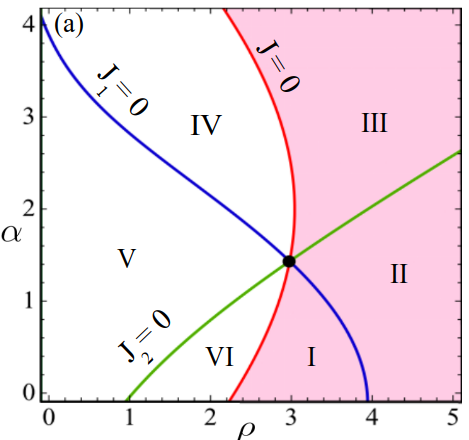}
\hspace{.2 cm}\includegraphics[height=5.08cm]
{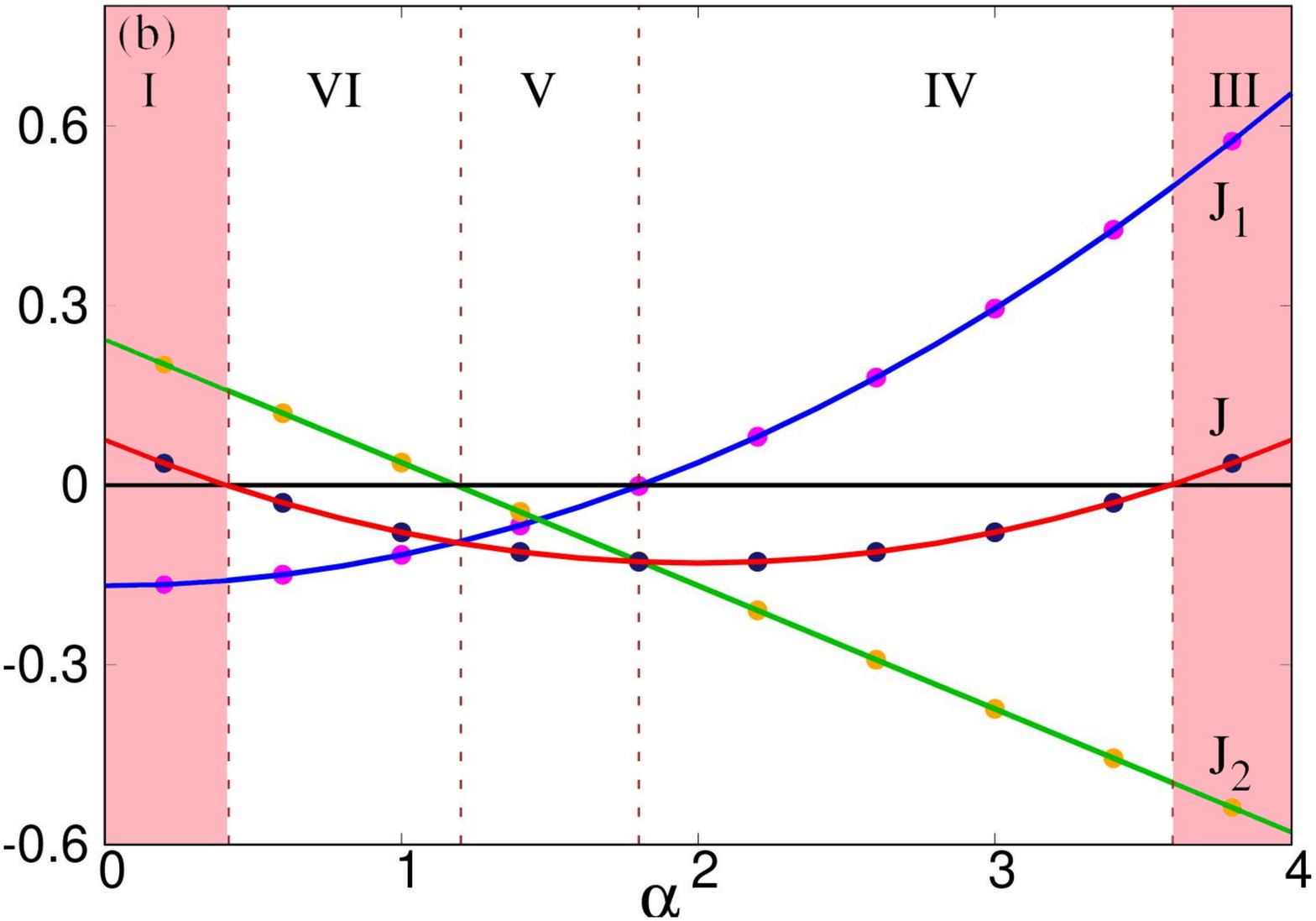} 
\caption{(a) Different regions in $\rho$ - $\alpha$ plane corresponding the direction of flow of the currents $J_{1,2}$ and $J$. Three lines $J=0$, $J_{1}=0$ and $J_{2}=0$ separate this plane in six regions. In the shaded regions I ($J_{1}<0$, $J_{2}>0$  and  $J_{2}>  |J_{1}|$), II ($J_{1}>0$ and  $J_{2}>0$) and III ($J_{1}>0$, $J_{2}<0$  and  $J_{1}>  |J_{2}|$), total current $J$ flows towards right. Similarly the current $J$ flows towards left in the regions IV ($J_{1}>0$, $J_{2}<0$ and $|J_{2}|> J_{1}$), V ($J_{1}<0$ and  $J_{2}<0$) and VI ($J_{1}<0$,  $J_{2}>0$ and $|J_{1}|> J_{2}$), (b) $J_{1,2}$ and $J$ as a function of $\alpha$ for   $\rho=2.5$. Five different regions are visible at $\rho=2.5$, the total current $J$ exhibits a {\it re-entrance}. For $\alpha <0.411$, $J$ is positive, the direction of $J$ is reversed with the increase of $\alpha$ (regions separated by dashed lines). Further for $\alpha > 3.589$, $J$ re-enters to the regime of forward flow. Regions marked here carry the same information as in (a). Points are from simulations with $L=100$ and averaged over $10^{8}$ trajectories, solid lines are exact according to  Eqs. (\ref{eq:ladder_fss_J1}) - (\ref{eq:ladder_fss_J}). }
 \label{fig:ladder_fss_J}
\end{figure}

One can show following Eqs. (\ref{eq:ladder_FSS_con_u}) and  (\ref{eq:ladder_FSS_con_v}) that this particle hopping model on a ladder has a FSS when the asymmetric rate functions satisfy the following distinct  functional forms 
\begin{eqnarray}
\fl u_{R}(n,m) = u(n,m)~ [1-\delta +\gamma u(n-1,m)] ~{\rm ;}~
u_{L}(n,m) = u(n,m)~ [\delta - \gamma u(n-1,m)] \label{eq:fn_u},\\
\fl v_{R}(n,m) = v(n,m)~ [\delta ' - \gamma ' v(n,m-1)] ~{\rm ;}~
v_{L}(n,m) = v(n,m)~ [1-\delta ' + \gamma ' v(n,m-1)] \label{eq:fn_v},\cr
\end{eqnarray}
characterized by  four independent  parameters $0\le \delta \le 1$,  $0 \le \gamma\le \delta/ u(n,m)|_{max}$ and $0\le \delta ' \le 1$,  $0 \le \gamma '\le \delta '/ v(n,m)|_{max}$. The range of $\delta$, $\gamma$ and $\delta '$, $\gamma '$ are chosen such that  the hop rates  $u_{R,L}(n)$ and  $v_{R,L}(n)$ remain   positive. We consider the rates $u(n,m)$ and $v(n,m)$ as 
\begin{eqnarray}
 u(n,m) = \frac{mn+n-m+1}{mn+n+2} ~~{\rm and}~~
 v(n,m) = \frac{mn+2}{mn+n+2} \label{eq:ladder_fss_uv},
\end{eqnarray}
which give a simple expression of the steady state weight  
\begin{equation}
f(n,m) =  \frac{mn+n+2}{2} \label{eq:ladder_fss_f}. 
\end{equation}
Let us consider the parameters, $\delta = \frac{n+1}{mn-m+n+1}$, $\gamma =\left( \frac{\alpha}{4}\right)^{2} \frac{m+n}{mn-2m+n}$ and $\delta ' = \frac{mn-n}{mn+2}$, $\gamma '=\left( \frac{\alpha-2}{4}\right) \frac{m+n}{mn-n+2}$, following Eqs. (\ref{eq:fn_u}) and (\ref{eq:fn_v}), we get 
\begin{eqnarray}
 u_{R}(n,m) = \frac{mn-m}{mn+n+2} + \left(\frac{\alpha}{4}\right)^{2} \frac{m+n}{mn+n+2}, \label{eq:ladder_fss_uR}\\
 u_{L}(n,m) = \frac{n+1}{mn+n+2} - \left(\frac{\alpha}{4}\right)^{2} \frac{m+n}{mn+n+2}, \label{eq:ladder_fss_uL}\\
 v_{R}(n,m) = \frac{mn-n}{mn+n+2} - \frac{(\alpha -2)(m+n)}{4 (mn+n+2)} \label{eq:ladder_fss_vR},\\
 v_{L}(n,m) = \frac{n+2}{mn+n+2} + \frac{(\alpha -2)(m+n)}{4 (mn+n+2)} \label{eq:ladder_fss_vL}.
\end{eqnarray}
The model parameter $\alpha$ has been taken in the range $[0,4]$ such that all the rates in Eqs.  (\ref{eq:ladder_fss_uR}) - (\ref{eq:ladder_fss_vL}) remain positive. We can express the grand canonical partition function following Eq. (\ref{eq:MB_ladder_QLN_FSS}),  $Z_{L}(z) = \sum_{N=0}^{\infty} z^{N} Q_{L,N} = [F(z)]^{L}$ with
\begin{equation}
 F(z) = \sum_{n=0}^{\infty} \sum_{m=0}^{\infty} z^{n} z^{m} f(n,m) = \frac{2z^{2}-3z+2}{2(z-1)^{4}},
\end{equation}
where the fugacity $z$ controls the particle density through the relation 
\begin{equation}
\rho(z)= (z/2) F'(z)/F(z) =\frac{z((5-4z)z-5)}{2(z-1)(2+z(2z-3))}.\label{eq:fss_rho_z}
\end{equation}
In a similar way, one can calculate the particle densities in both legs, $\rho_{1}$ in the lower leg and $\rho_{2}$ in the upper leg as 
\begin{equation}
 \rho_{1}(z) =\frac{z(3z-2z^{2}-3)}{(z-1)(2+z(2z-3))} ~~{\rm and}~~ 
\rho_{2}(z)=\frac{2z(z-z^{2}-1)}{(z-1)(2+z(2z-3))}.
\label{eq:fss_rho1_rho2}
\end{equation}
Note that the densities $\rho$, $\rho_{1}$ and $\rho_{2}$  in Eqs. (\ref{eq:fss_rho_z}) and (\ref{eq:fss_rho1_rho2}) do not depend on $\alpha$; {\it i.e,} for any value of $\alpha$, a given $z$ corresponds to a unique density $\rho$. At $z\rightarrow 0$ ({\it i.e,} $\rho \rightarrow 0$), both the densities $\rho_{1} \rightarrow 0$ and $\rho_{2} \rightarrow 0$ but their ratio remains finite, $\frac{3}{2}$. Similarly for $z\rightarrow 1$ ({\it i.e,} $\rho \rightarrow \infty$), both $\rho_{1} \rightarrow \infty$ and $\rho_{2} \rightarrow \infty$ but the ratio becomes $1$. The relative particle density $\frac{\rho_{1}}{\rho_{2}}$ as a function of the total density of the system $\rho$ has been shown in Fig. \ref{fig:ladder_fss_rho1_rho2}(a).
We can calculate the currents in both legs, $J_{1}$ for lower leg and $J_{2}$ for upper leg as
\begin{eqnarray}
\fl J_{1} =\frac{1}{F(z)} \sum_{n=1}^{\infty} \sum_{m=0}^{\infty}~ [u_{r}(n,m)-u_{l}(n,m)]~ z^{n+m} f(n,m)  = \frac{z(z(24-z\alpha^{2})+\alpha^{2}-16)}{4(4z^{2}-6z+4)}, \label{eq:ladder_fss_J1}\\
\fl J_{2} =\frac{1}{F(z)} \sum_{n=0}^{\infty} \sum_{m=1}^{\infty} ~[v_{r}(n,m)-v_{l}(n,m)]~ z^{n+m} f(n,m)  = \frac{z(z(6+z(\alpha-2))-\alpha-2)}{4z^{2}-6z+4}\cr \label{eq:ladder_fss_J2}
\end{eqnarray}
and the total current can be expressed as
\begin{equation}
 J=J_{1}+J_{2}=\frac{z((\alpha -2)^{2}(1-z^{2})-4(z^{2}-12z+7))}{4(4z^{2}-6z+4)}. \label{eq:ladder_fss_J}
\end{equation}
For $z\rightarrow 1$,  the density $\rho \rightarrow \infty$ and the total current $J$ turns out to be $J=2$. For $z\rightarrow 0$, {\it i.e,} when $\rho \rightarrow 0$, the total current $J$ vanishes. Infact there exists a line in $\rho$ - $\alpha$ parameter space, where the total current $J$ also vanishes. This line can be expressed following Eq. (\ref{eq:ladder_fss_J}) as
\begin{equation}
 z^{*}=\frac{24-\sqrt{(\alpha-2)^{4}-24(\alpha-2)^{2}+464}}{(\alpha-2)^{2}+4}\label{eq:fss_J0line}
\end{equation}
where, $z^{*}$ is the fugacity. Since $z^{*}$ corresponds to a unique density $\rho^{*} = \rho(z^{*})$, $J=0$ line can also be expressed in terms of the density $\rho^{*}$ as a function of $\alpha$ as
\begin{equation}
\rho^{*}=1-\frac{\kappa^{2}}{8}+\frac{\sqrt{\kappa^{4}-24\kappa^{2}+464}}{8} + \frac{16(\kappa^{2}-12)-96\sqrt{\kappa^{4}-24\kappa^{2}+464}}{3312+7\kappa^{2}(\kappa^{2}-24)}\label{eq:fss_J0line_rho}
\end{equation}
where $\kappa=(\alpha-2)$. Note that $\rho^{*}$ corresponds to the value of density at which the total current $J$ reverses its direction. $\rho^{*}$ as a function of $\alpha$ is shown in Fig. \ref{fig:ladder_fss_J}(a) and marked as $J=0$ line (the red one). As expected, $\rho^{*}$ as a function of $\alpha$ is symmetric about $\alpha=2$ and has a maximum  at $\alpha=2$. Corresponding densities for $\alpha=0$ and $4$ are $\rho_{1}^{*}=(\frac{39}{86}+\frac{31\sqrt{6}}{43}) \approx 2.219$ and at $\alpha=2$ the density becomes $\rho_{2}^{*}=(\frac{65}{69}+\frac{53\sqrt{29}}{138}) \approx 3.01$. For $\alpha=0.6$, the total current $J$ reverses its direction, following Eq. (\ref{eq:fss_J0line_rho}), at $\rho=2.611$ which is shown in Fig. \ref{fig:ladder_fss_rho1_rho2}(b).
In the $\rho$ - $\alpha$ plane, we have shown the  three lines of separations, $J_1 =0, J_2=0, J=0;$   corresponding currents flip
their direction   when these lines are crossed by  varying  the density $\rho$ or the parameter $\alpha.$  In the shaded region 
of  Fig. \ref{fig:ladder_fss_J}(a),  the  total current $J$ flows  towards right.  This is possible  when  I. $J_{1}<0$, $J_{2}>0$  and  $J_{2}>  |J_{1}|$  or II. when both  $J_{1}>0$ and  $J_{2}>0$  or III.  when  $J_{1}>0$, $J_{2}<0$  and  $J_{1}>  |J_{2}|.$
Similarly  when  total current $J$ flows  towards left,  we have 
three more regions  IV. $J_{1}>0$, $J_{2}<0$ and $|J_{2}|> J_{1}$ , V. both  $J_{1}<0$ and  $J_{2}<0$ and VI.  $J_{1}<0$,  $J_{2}>0$ and $|J_{1}|> J_{2}$.    All these regions are marked in  
Fig. \ref{fig:ladder_fss_J}(a).   It is evident from the figure 
that one can access  at most four regions  by changing $\rho$ for a fixed $\alpha$, whereas at most five regions can be accessed  by changing $\alpha$ for a fixed $\rho$. The total current $J$, when $\rho$ is varied, does not exhibit re-entrance, whereas it shows re-entrance when $\alpha$ is varied in a certain zone. This re-entrance of the total current $J$ as a function of $\alpha$ occurs in the density region $\rho_{1}^{*} < \rho < \rho_{2}^{*}$. 
In Fig. \ref{fig:ladder_fss_J}(b)  we have plotted $J_{1,2}$ and $J$  as a function of $\alpha$ keeping the particle density fixed at $\rho=2.5;$  five different regions are clearly visible. At $\alpha=0$, $J_{1}$ is negative, $J_{2}$ is positive and $J_{2}>|J_{1}|$. Thus at $\alpha=0$ and for small $\alpha$, the total current $J=J_{1}+J_{2}$ is positive. Note that, $\frac{dJ_{1}}{d\alpha}>0$ and $\frac{dJ_{2}}{d\alpha}<0$ following Eqs. (\ref{eq:ladder_fss_J1}) and (\ref{eq:ladder_fss_J2}). Since, $J_{1}$ increases faster  than $J_{2}$, it is evident that for larger $\alpha$ we may get the total current $J$ positive again. Infact, the total current $J$ vanishes  at $\alpha=0.411$   and  then  the direction of the total current is reversed.  Further increase  of  $\alpha$  keeps the direction of the total current unaltered until  $\alpha=3.589,$ where  $J$ vanishes  and then reverses  its direction again, re-entering to  regime of forward-flow and exhibits the {\it re-entrance}. 

\section{Two-leg ladder with pair  factorized steady state \label{sec:PFSS}}

When the steady state is factorized as a product of two site clusters, it is commonly known as pair factorized steady state (PFSS). It was proposed and studied in \cite{pfss}, where it was shown that for a particular class of PFSS, the system under consideration exhibits a condensation transition. Later PFSS has been found  in continuous mass-transfer models \cite{mass_Bertin, mass_waclaw}, in systems with open boundaries \cite{open_pfss} and in random graphs \cite{graph_waclaw}, etc. In our ladder example, we assume that the system evolves to the PFSS which looks like 
\begin{eqnarray} 
\fl P(\{n_{i},m_{i}\}) = \frac{1}{Q_{L,N}}\prod_{i=1}^{L} 
 g(n_{i},m_{i}, n_{i+1},m_{i+1} )
~\delta \left(\sum_{i} (n_{i}+m_{i}) - N\right)  \label{eq:MB_ladder_P_PFSS}
\end{eqnarray}
with a canonical partition function
\begin{equation}\label{eq:MB_ladder_QLN_PFSS}
Q_{L,N} =\sum_{\{n_{i}\};\{m_{i}\}}\prod_{i=1}^{L} 
  g(n_{i},m_{i}, n_{i+1},m_{i+1} )
~\delta \left(\sum_{i} (n_{i}+m_{i}) - N\right).
\end{equation}
If the hop rate depends only on the occupation number of the departure site  and the corresponding  site on the other leg,  PFSS as in Eq. (\ref{eq:MB_ladder_P_PFSS}) is not possible in general. We  generalize that the  hop rates now  depend not only on the occupation number of departure site, also on the occupation numbers of its two nearest neighbours on both legs, {\it i.e,} $u\equiv u(n_{i-1},m_{i-1},n_{i},m_{i},n_{i+1},m_{i+1})$. Here also we ask if such a pair factorized form of steady state is possible when all the rate functions are different. We, using  similar arguments of MB as described for FSS, find that 
a pair factorized form of steady state (PFSS) as in Eq. (\ref{eq:MB_ladder_P_PFSS})  is possible   for the  two-leg ladder when the hop rates satisfy
\begin{eqnarray}
\fl [~ u_{R}(n_{i-1},m_{i-1},n_{i},m_{i},n_{i+1},m_{i+1}) + u_{L}(n_{i-1},m_{i-1},n_{i},m_{i},n_{i+1},m_{i+1})~]\cr \fl =u(n_{i-1},m_{i-1},n_{i},m_{i},n_{i+1},m_{i+1}) = \frac{g(n_{i-1},m_{i-1},n_{i}-1,m_{i})}{g(n_{i-1},m_{i-1},n_{i},m_{i})} \frac{g(n_{i}-1,m_{i},n_{i+1},m_{i+1})}{g(n_{i},m_{i},n_{i+1},m_{i+1})},\cr \label{eq:ladder_PFSS_con_u} \\
\fl [~ v_{R}(n_{i-1},m_{i-1},n_{i},m_{i},n_{i+1},m_{i+1}) + v_{L}(n_{i-1},m_{i-1},n_{i},m_{i},n_{i+1},m_{i+1})~]\cr \fl =v(n_{i-1},m_{i-1},n_{i},m_{i},n_{i+1},m_{i+1}) = \frac{g(n_{i-1},m_{i-1},n_{i},m_{i}-1)}{g(n_{i-1},m_{i-1},n_{i},m_{i})} \frac{g(n_{i},m_{i}-1,n_{i+1},m_{i+1})}{g(n_{i},m_{i},n_{i+1},m_{i+1})}.\cr
 \label{eq:ladder_PFSS_con_v}
\end{eqnarray} 
Let us consider that the weight function $g(n_{i},n_{i+1},m_{i},m_{i+1})$ can be written as the inner product of two 2-dimensional vectors \cite{FRP}
\begin{equation}\label{eq:MB_ladder_gmn}
 g(n_{i},n_{i+1},m_{i},m_{i+1}) = \langle \alpha (n_{i},m_{i}) | \beta (n_{i+1},m_{i+1}) \rangle
\end{equation}
In the grand canonical ensemble where the fugacity $z$ controls the density $\rho$, the partition sum can be written as 
$Z_{L}(z)= \sum_{N=0}^{\infty} Q_{L,N}z^{N}
= \Tr[T(z)^{L}]$ with
\begin{equation}\label{eq:MB_ladder_T}
 T(z) = \sum_{n=0}^{\infty} \sum_{m=0}^{\infty} z^{n} z^{m} |\beta (n,m) \rangle \langle \alpha (n,m)|.
\end{equation}
We consider a simple 2-dimensional representation by taking,
\begin{eqnarray}
 \langle \alpha (n,m)| = ((n+1)^{-\nu}(m+1)^{-\nu} , (m+1)^{-\nu}), \cr \langle \beta (n,m)| = ((n+1)^{1-\nu},(n+1)^{1-\nu}(m+1)^{1-\nu}). \label{eq:ladder_pfss_rep}
\end{eqnarray}
The weight function $g(n_{i},n_{i+1},m_{i},m_{i+1})$ can be determined for the above choice of representation  
following Eq. (\ref{eq:MB_ladder_gmn}). The transfer matrix $T(z)$, following Eq. (\ref{eq:MB_ladder_T}), becomes 
\begin{equation} \label{eq:ladder_pfss_T}
T(z) = \frac{1}{z^{2}} \left(
\begin{array}{cc}
Li_{\nu}(z) Li_{2\nu -1}(z) & Li_{\nu -1}(z)Li_{\nu}(z) \\
(Li_{2\nu -1}(z))^{2} & Li_{\nu -1}(z)Li_{2\nu -1}(z)
\end{array}
\right),
\end{equation}
where $Li_{\nu}(z)$ is the Polylogarithm function defined by $Li_{\nu}(z) = \sum_{n=1}^{\infty} \frac{z^{n}}{n^{\nu}}$. The eigenvalues of $T(z)$ are 
\begin{equation}
 \lambda_{+} =\frac{1}{z^{2}}(Li_{\nu}(z)+Li_{\nu -1}(z))Li_{2\nu -1}(z) ~~{\rm and}~~ \lambda_{-}=0.
\end{equation}
The partition function $Z_{L}(z)$ in the thermodynamic limit becomes $Z_{L}(z) \simeq \lambda_{+}(z)^{L} $ which leads to the density fugacity relation 
\begin{equation}
\rho(z)=(z/2)\frac{\partial }{\partial z}\ln (\lambda_{+}(z)) = \frac{Li_{\nu-2}(z)+Li_{\nu-1}(z)}{2(Li_{\nu-1}(z)+Li_{\nu}(z))}+\frac{Li_{2\nu-2}(z)}{2Li_{2\nu-1}(z)}-1
\end{equation}
and the critical density $\rho_{c}=  \lim_{z \rightarrow 1} \rho(z)$. 
The hop rates $u(n_{i-1},m_{i-1},n_{i},m_{i},n_{i+1},m_{i+1})$ and $v(n_{i-1},m_{i-1},n_{i},m_{i},n_{i+1},m_{i+1})$ can be determined following  Eq. (\ref{eq:ladder_pfss_rep}) as
\begin{eqnarray}
\fl  u(n_{i-1},m_{i-1},n_{i},m_{i},n_{i+1},m_{i+1}) = \left(\frac{n_{i}}{n_{i}+1}\right)^{1-2\nu}\frac{n_{i}^{\nu}(1+m_{i+1})+(1+m_{i+1})^{\nu}}{(1+m_{i+1})^{\nu}+(1+m_{i+1})(1+n_{i})^{\nu}}, \label{eq:ladder_pfss_u} \\
\fl  v(n_{i-1},m_{i-1},n_{i},m_{i},n_{i+1},m_{i+1}) = \left(\frac{m_{i}}{m_{i}+1}\right)^{-2\nu}\frac{m_{i}^{\nu}+m_{i}(1+n_{i-1})^{\nu}}{(1+m_{i})^{\nu}+(1+m_{i})(1+n_{i-1})^{\nu}}. \label{eq:ladder_pfss_v}
\end{eqnarray}
We consider the rates in the horizontal directions
\begin{eqnarray}
\fl u_{R,L}(n_{i-1},m_{i-1},n_{i},m_{i},n_{i+1},m_{i+1}) = (u(n_{i-1},m_{i-1},n_{i},m_{i},n_{i+1},m_{i+1}) \pm (\alpha ^{2}/2-1))/2, \cr \label{eq:ladder_pfss_uRL}\\
\fl v_{R,L}(n_{i-1},m_{i-1},n_{i},m_{i},n_{i+1},m_{i+1}) = (v(n_{i-1},m_{i-1},n_{i},m_{i},n_{i+1},m_{i+1}) \pm (1-\alpha))/2, \label{eq:ladder_pfss_vRL}
\end{eqnarray}
where in Eqs. (\ref{eq:ladder_pfss_uRL}) and (\ref{eq:ladder_pfss_vRL}), $u_{R}$ and $v_{R}$ corresponds to $'+'$ sign and $u_{L}$ and $v_{L}$ corresponds to $'-'$ sign, such that they satisfy the conditions in Eqs. (\ref{eq:ladder_PFSS_con_u}) and (\ref{eq:ladder_PFSS_con_v}). One can verify, following Eqs. (\ref{eq:ladder_PFSS_con_u}) and (\ref{eq:ladder_PFSS_con_v}) that the average hop rates along the vertical direction become equal for both lower and upper leg as  
\begin{eqnarray}
\fl \langle  u(n_{i-1},m_{i-1},n_{i},m_{i},n_{i+1},m_{i+1}) \rangle = \langle  v(n_{i-1},m_{i-1},n_{i},m_{i},n_{i+1},m_{i+1}) \rangle = z.
\end{eqnarray}
We can explicitly calculate the current $J_{1}$ in the lower leg and $J_{2}$ in the upper leg as
\begin{eqnarray}
 J_{1} =\langle u_{R}(.) - u_{L}(.) \rangle = (\alpha ^{2}/2-1) ~\left(1-\frac{Tr[~T(z)^{L-1}~T_{1}(z)~]}{Tr(T(z))^{L}} \right) \cr 
 ~~~~= (\alpha ^{2}/2-1)\left(1-\frac{z~[~Li_{\nu}(z)+Li_{2\nu-1}(z)~]}{[~Li_{\nu}(z)+Li_{\nu-1}(z)~]~Li_{2\nu-1}(z)} \right),
 \label{eq:ladder_pfss_J1} \\
 J_{2}=\langle v_{R}(.) - v_{L}(.) \rangle = (1-\alpha) ~\left(1-\frac{Tr[~T(z)^{L-1}~T_{2}(z)~]}{Tr(T(z))^{L}} \right) \cr 
  ~~~~=(1- \alpha) \left(1-\frac{z~[~Li_{\nu-1}(z)+Li_{2\nu-1}(z)~]}{[~Li_{\nu}(z)+Li_{\nu-1}(z)~]~Li_{2\nu-1}(z)} \right),\label{eq:ladder_pfss_J2}
\end{eqnarray}
where the matrices
\begin{equation} 
\fl T_{1}(z) = \frac{1}{z} \left(
\begin{array}{cc}
Li_{\nu}(z)  & Li_{\nu}(z) \\
Li_{2\nu -1}(z) & Li_{2\nu -1}(z)
\end{array}
\right) ~~{\rm and}~~
T_{2}(z) = \frac{1}{z^{}} \left(
\begin{array}{cc}
Li_{2\nu -1}(z) & Li_{\nu -1}(z) \\
Li_{2\nu -1}(z) & Li_{\nu -1}(z)
\end{array}
\right).
\end{equation}
The total current of the system is  $J=J_{1}+J_{2}.$
\begin{figure}[h]
\vspace*{.5 cm}
\centering
 \includegraphics[height=4.9cm]{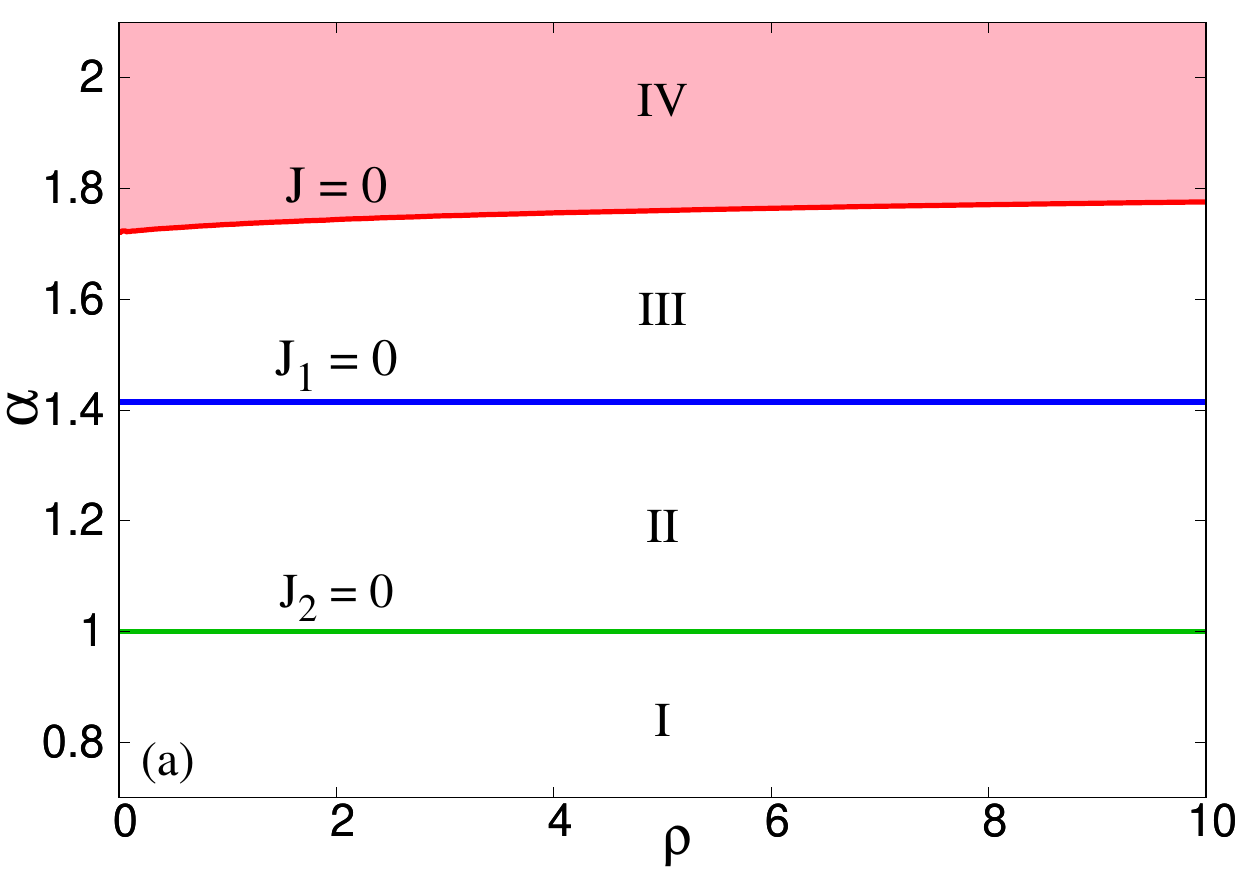}
 \hspace{.2 cm}\includegraphics[height=5.1 cm]{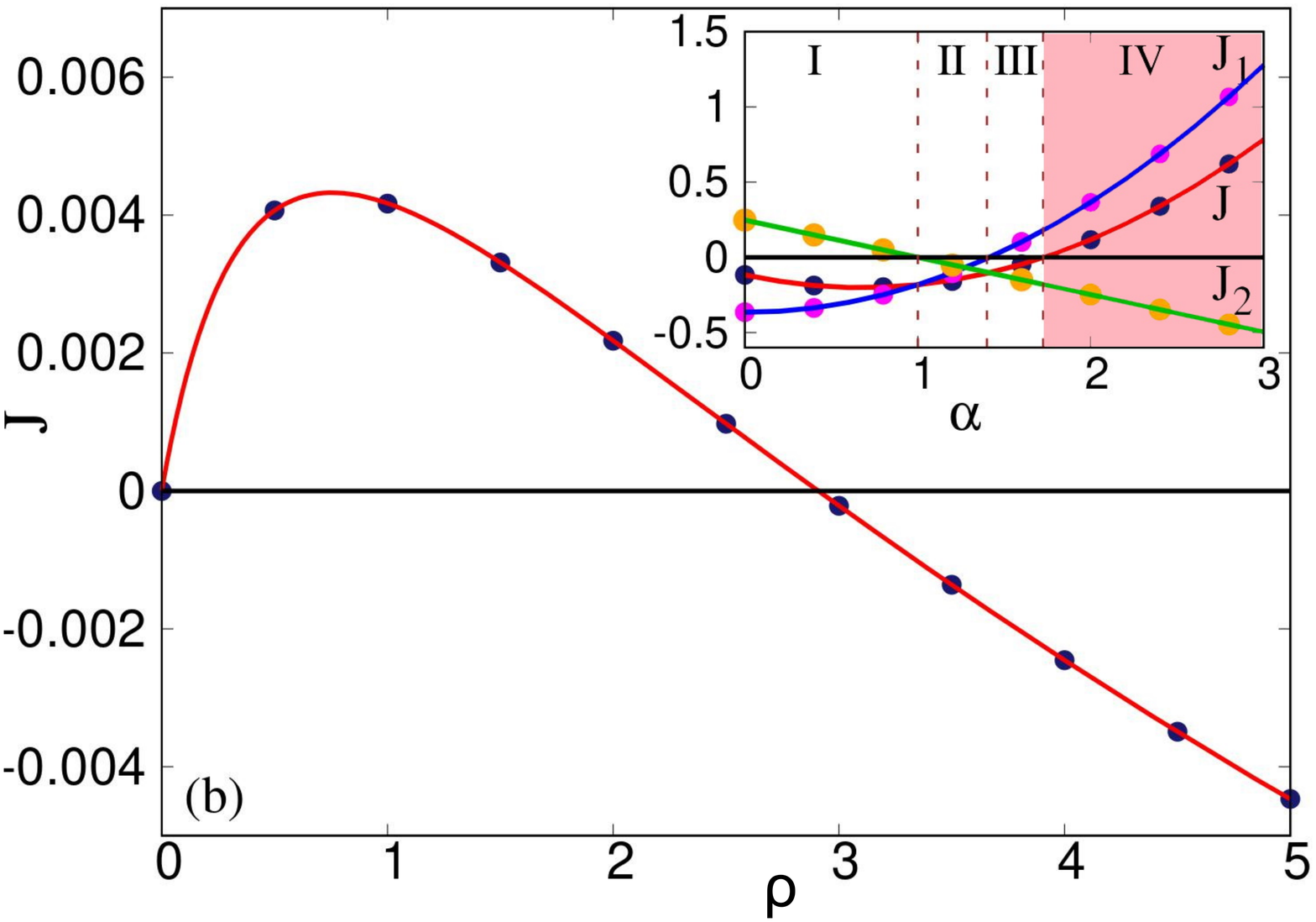}
\caption{(a) Different regions in $\rho$ - $\alpha$ plane corresponding the direction of flow of the currents $J_{1,2}$ and $J$ for $\nu=1$. Three lines $J=0$, $J_{1}=0$ and $J_{2}=0$ separate this plane in four regions. In the shaded region IV ($J_{1}>0$, $J_{2}<0$  and  $J_{1}>  |J_{2}|$), the total current $J$ flows towards right. Similarly the total current flows towards left in the regions I ($J_{1}<0$,  $J_{2}>0$ and $|J_{1}|> J_{2}$),  II ($J_{1}<0$ and  $J_{2}<0$) and III ($J_{1}>0$, $J_{2}<0$ and $|J_{2}|> J_{1}$). (b) The total current $J$ as a function of density $\rho$ for $\alpha=1.75$. The total current reverses its direction at density $\rho=2.923$. Inset shows $J_{1,2}$ and $J$ as a function of $\alpha$ for $\nu=1$ and  $\rho=0.5$. All four  regions are visible at $\rho=0.5$. For small $\alpha$ , $J<0$ and the direction is reversed for $\alpha >1.725$.
Regions, separated by dashed lines, carry the same information as in figure (a). Points are from simulations with $L=100$ and averaged over $10^{8}$ trajectories, solid lines are exact according to Eqs. (\ref{eq:ladder_pfss_J1}) - (\ref{eq:ladder_pfss_J2}).}
\label{fig:ladder_pfss_J}
\end{figure}
We consider a particular case, when $\nu=1$. 
In the $\rho$ - $\alpha$ plane, we have shown the  three lines of separations, $J_1 =0, J_2=0, J=0;$   corresponding currents flip
their direction   when these lines are crossed by  varying  the density $\rho$ or the parameter $\alpha.$  In the shaded region (IV)
of  Fig \ref{fig:ladder_pfss_J} (a),  the  total current $J$ flows  towards right. For our choices of rates, this is possible only when $J_{1}>0$, $J_{2}<0$  and  $J_{1}>  |J_{2}|.$
Similarly  when the total current $J$ flows  towards left,  we have three more regions  I. $J_{1}<0$,  $J_{2}>0$ and $|J_{1}|> J_{2}$, II. both  $J_{1}<0$ and  $J_{2}<0$ and III. $J_{1}>0$, $J_{2}<0$ and $|J_{2}|> J_{1}$. All these regions are marked in  
Fig. \ref{fig:ladder_pfss_J}(a). One can access  at most two regions  by changing $\rho$ for a fixed $\alpha$ (region IV to III), whereas  all four regions can be accessed by   changing $\alpha$ for a fixed $\rho.$ It is evident from Fig. \ref{fig:ladder_pfss_J}(a) that the total current $J$ as a function of density $\rho$ reverses its direction in a certain zone. This current reversal occurs in the region $1.72<\alpha \leq 1.78$. For $\alpha=1.75$, direction of the total current $J$ is reversed at $\rho \approx 2.923$, which is shown in Fig. \ref{fig:ladder_pfss_J}(b).  In the inset of Fig. \ref{fig:ladder_pfss_J}(b), $J_{1,2}$ and $J$ have been plotted as a function of $\alpha$  for a fixed particle density at $\rho=0.5;$  all four regions are clearly visible here. For small $\alpha$, $J$ is negative and vanishes at $\alpha=1.725$ then the direction of the current is reversed. Further increase of $\alpha$ keeps $J$ in this same  direction.

\section{Multibalance in other nonequilibrium lattice models\label{sec:CFSS}}

Multibalance condition that we introduced is very useful in solving nonequilibrium problems. A few examples are given below.
\subsection{Asymmetric finite range process (AFRP) with nearest neighbour and next nearest neighbour hopping}

\begin{figure}[h]
\vspace*{.5 cm}
\begin{center}
\includegraphics[scale = 0.3]{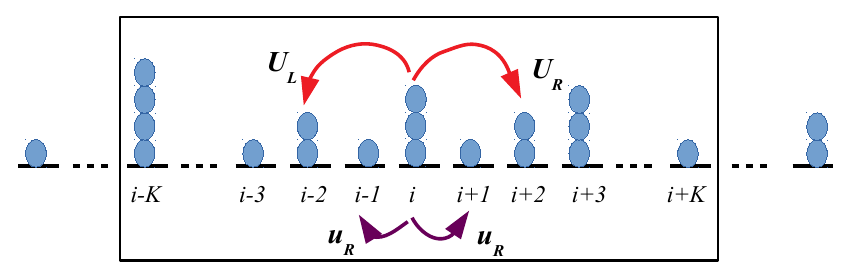}
\end{center}
\caption{ FRP in one dimension where one particle hops from a site $i$ to its left and right neighbours with rates $u_{L}(.)$  and $u_{R}(.)$ and left and right next nearest neighbours with  rates $U_{L}(.)$ and $U_{R}(.)$. All these rates  depend on occupation of site $i$ (here $n_{i} = 3$) and all its neighbours within a range $K$.}
\label{fig:MB_FRP_pic}
\end{figure}

Let us consider a system of $N$ particles on a one dimensional periodic lattice with $L$ sites labeled by $i = 1,2,\cdots ,L$ (see Fig. \ref{fig:MB_FRP_pic}). Each site $i$ can accommodate  $n_{i}\geq 0$ number of particles. From a randomly chosen site $i$, one particle can hop to its right nearest neighbour with rate $u_{R}(.)$, left  nearest neighbour with rate $u_{L}(.)$ and as well as to the next nearest neighbours with rates $U_{R}(.)$ for right, $U_{L}(.)$ for left. All these rates depend on the number of particles at all the sites within a range $K$ w.r.t the departure site \cite{FRP, asymm_FRP}.
We assume that the system evolves to a cluster factorized steady state
\begin{equation} \label{eq:MB_FRP_prob}
 P(\{n_{i}\}) \propto \prod_{i=1}^{L} 
 g(n_{i}, n_{i+1}, \cdots n_{i+K})
 ~\delta \left(\sum_{i} n_{i} - N\right).
\end{equation}
$N$ is the total number of particles and $\rho = \frac{N}{L}$ is conserved by the dynamics. We now ask, if such a cluster factorized form of steady state is possible when the rates in all four directions are different {\it i.e} the ratios of rates are not independent of $n$. We can say that steady state as in Eq. (\ref{eq:MB_FRP_prob}) can be obtained using MB for a given function $u(n_{i-K}, \cdots , n_{i}, \cdots n_{i+K})$, when the hop rates satisfy $u_{R}(n_{i-K}, \cdots , n_{i}, \cdots n_{i+K}) =u_{L}(n_{i-K}, \cdots , n_{i}, \cdots n_{i+K})=  u(n_{i-K}, \cdots , n_{i}, \cdots n_{i+K}) $ and 
\begin{eqnarray}
U_{R}(n_{i-K}, \cdots , n_{i}, \cdots n_{i+K}) +U_{L}(n_{i-K}, \cdots , n_{i}, \cdots n_{i+K}) \cr = u(n_{i-K}, \cdots , n_{i}, \cdots n_{i+K})  = \prod_{k=1}^{K}
  \frac{g(\widetilde n_{i-K+k}, \widetilde n_{i-K+1+k},\cdots , \widetilde n_{i+k} )}{g(n_{i-K+k},n_{i-K+1+k}, \cdots , n_{i+k} )},\label{eq:MB_FRP_cond}
 \end{eqnarray}
where $\widetilde n_{j} = n_{j} - \delta_{ji}$.

\subsection{Two species FRP  with directional asymmetry}

\begin{figure}[h]
\centering
\includegraphics[scale = 0.3]{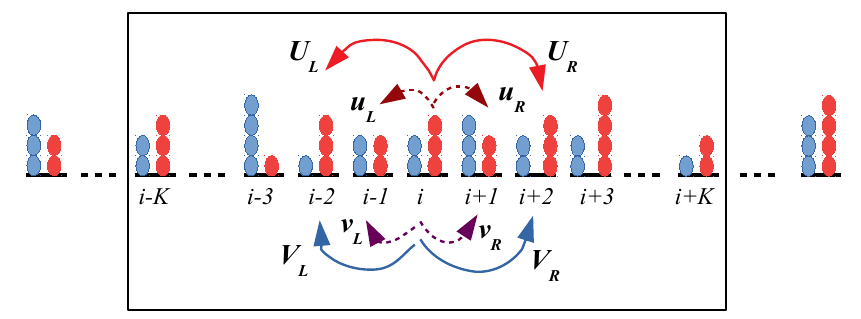}
\caption{Two species asymmetric FRP  model in one dimension. A particle of species A (coloured red) can hop to its right and left  nearest neighbours with rates $u_{R}(.)$ and $u_{L}(.)$, to right and left next nearest neighbours with rates $U_{R}(.)$ and $U_{L}(.)$. Similarly, particle of species $B$ (coloured blue), can hop to its right and left nearest neighbours with rates $v_{R}(.)$ and $v_{L}(.)$, to right and left next nearest neighbours with rates $V_{R}(.)$ and $V_{L}(.)$.}
 \label{fig:MB_frp_model}
\end{figure}

We consider a one dimensional periodic lattice
(see Fig. \ref{fig:MB_frp_model}) with sites labeled by $i = 1,2, \cdots , L$. At each site $i$, there are $n_{i}$ particles of species $A$ (coloured red) and $m_{i}$ particles of species $B$ (coloured blue). From a  randomly chosen site $i$, a particle of species $A$, can hop to its right and left nearest neighbours with rates $u_{R}(.)$ and $u_{L}(.)$, it can hop to right and left next nearest neighbours with rates $U_{R}(.)$ and $U_{L}(.)$. 
Similarly, a particle of species $B$, can hop to its right and left nearest neighbours with rates $v_{R}(.)$ and $v_{L}(.)$, to right and left next nearest neighbours with rates $V_{R}(.)$ and $V_{L}(.)$. All these rates  depend on the number  of particles of both species at the departure site and its two nearest neighbours. 
We demand that this model evolves to a PFSS
\begin{eqnarray} 
\fl P(\{n_{i},m_{i}\}) \propto \prod_{i=1}^{L} 
 g(n_{i}, m_{i}, n_{i+1},m_{i+1} )
~\delta \left(\sum_{i} n_{i} - N\right) \delta \left(\sum_{i} m_{i} - M\right), \label{eq:MB_2FRP_P_PFSS}
\end{eqnarray}
where, $N$ is total number of particles of species $A$ and $M$ is total number of particles of species $B$. Such a PFSS as in Eq. (\ref{eq:MB_2FRP_P_PFSS}) can be obtained using MB for the given functions of $u(n_{i-1},m_{i-1},n_{i},m_{i},n_{i+1},m_{i+1})$ and $v(n_{i-1},m_{i-1},n_{i},m_{i},n_{i+1},m_{i+1})$, when  the hop rates of species $A$ satisfy $u_{R}(n_{i-1},m_{i-1},n_{i},m_{i},n_{i+1},m_{i+1}) = u_{L}(n_{i-1},m_{i-1},n_{i},m_{i},n_{i+1},m_{i+1}) = u(n_{i-1},m_{i-1},n_{i},m_{i},n_{i+1},m_{i+1})$ and 
\begin{eqnarray} 
\fl \left[~U_{R}(n_{i-1},m_{i-1},n_{i},m_{i},n_{i+1},m_{i+1}) + U_{L}(n_{i-1},m_{i-1},n_{i},m_{i},n_{i+1},m_{i+1})~ \right] \cr
\fl = 
  u(n_{i-1},m_{i-1},n_{i},m_{i},n_{i+1},m_{i+1})
 = \frac{g(n_{i-1},m_{i-1},n_{i}-1,m_{i})}{g(n_{i-1},m_{i-1},n_{i},m_{i})} \frac{g(n_{i}-1,m_{i},n_{i+1},m_{i+1})}{g(n_{i},m_{i},n_{i+1},m_{i+1})},\cr \label{eq:MB_2FRP_con1_PFSS}
\end{eqnarray} 
similarly, the hop rates of species $B$ satisfy  $v_{R}(n_{i-1},m_{i-1},n_{i},m_{i},n_{i+1},m_{i+1}) = v_{L}(n_{i-1},m_{i-1},n_{i},m_{i},n_{i+1},m_{i+1}) = v(n_{i-1},m_{i-1},n_{i},m_{i},n_{i+1},m_{i+1})$ and 
\begin{eqnarray} 
\fl \left[~V_{R}(n_{i-1},m_{i-1},n_{i},m_{i},n_{i+1},m_{i+1}) + V_{L}(n_{i-1},m_{i-1},n_{i},m_{i},n_{i+1},m_{i+1})~ \right] \cr  \fl = 
  v(n_{i-1},m_{i-1},n_{i},m_{i},n_{i+1},m_{i+1})
 =\frac{g(n_{i-1},m_{i-1},n_{i},m_{i}-1)}{g(n_{i-1},m_{i-1},n_{i},m_{i})} \frac{g(n_{i},m_{i}-1,n_{i+1},m_{i+1})}{g(n_{i},m_{i},n_{i+1},m_{i+1})}.\cr \label{eq:MB_2FRP_con2_PFSS}
\end{eqnarray}

\section{Summary}

The steady states of non-equilibrium systems are very much dependent on the complexity of the dynamics and  it is
difficult to track down a  systematic procedure to obtain the steady state measure of a system with a given dynamics. In this regard, starting from the master equation
that governs the time evolution of a many particle system in the configuration space,
several flux cancellation schemes have been in use  for obtaining  the exact  steady state  weight.  These schemes include matrix product ansatz (MPA) \cite{derrida__tasep_mpa}, h-balance scheme \cite{asymm_FRP} and pairwise balance condition (PWB). In this article  we introduced  a new kind of balance condition, namely multibalance (MB), where the sum of  incoming fluxes from  a set of configurations to any configuration $C$  is balanced by the total outgoing flux  to  set of configurations chosen suitably.

We have applied the MB condition to a class of nonequilibrium lattice models. We have given an example of the asymmetric  ZRP in  two dimensions and discussed that a factorized steady state (FSS) can be obtained using MB with specific conditions on hop rates. We have considered an interesting model, particle hopping on a ladder, where a particle from  a randomly chosen site in one leg, can hop to its two  nearest neighbours and also to that site of the other leg. We first  assumed that the model evolves to FSS and discussed the cases of obtaining FSS using detailed balance (DB) and PWB, which are well known. We ask that such a FSS is possible at all when hop rates in all directions are different. In this situation, MB can be employed to solve this model exactly to obtain the FSS with specific conditions on the hop rates.
We have calculated currents generated by both legs and also the total current for this model. It is shown that the total current flows in same direction in two different regions separated by an intermediate region, where it flows in the opposite direction. To explain this behaviour, we  have mentioned it clearly that total current being  positive or negative does not always mean that currents on  both legs are individually positive or negative. 

We extended  the  model of  particle hopping  on ladder beyond FSS; we have discussed that pair factorized form of steady state  (PFSS) can also be obtained using MB if the hop rates satisfy some other conditions. 

The problem which we studied here, can also be mapped to  hopping of particles on a periodic lattice, where each particle  has  two internal degrees of freedom $\sigma=\pm$ that replace the  two legs  of the ladder.  Then   $(n_{i}, m_{i})$   translates to   $(n^+_{i}, n^-_{i}).$ In this model, a particle from site $i$ can hop to its right and left nearest neighbour with rate $u_{R,\sigma}(n^+_{i}, n^-_{i})$ and $u_{L,\sigma}(n^+_{i}, n^-_{i})$  or   it  can  change  it's  internal degree of freedom $\sigma  \to - \sigma$ with  rate $u_{\sigma}(n^+_{i}, n^-_{i})$, which violates the conservation of particles of each kind. $J_{\pm}\equiv J_{1,2}$  are the currents generated by each kind of particles and $J$ is the sum  of $J_+$ and $J_-.$

To emphasize the utility and the strength of the MB condition introduced here, we  have  discussed a few examples of nonequilibrium lattice models where we can obtain steady state using MB. For the asymmetric finite range process, with nearest and next nearest neighbours hopping, cluster factorized form of steady state (CFSS) can be obtained using MB for certain conditions on hop rates, which helps us calculating the steady state average of  
the observable using Transfer Matrix method  introduced  earlier \cite{FRP}. In another interesting example, we have considered a  two species FRP with directional asymmetry, and with nearest neighbours and next nearest neighbours hopping. PFSS can be obtained for this model under specific conditions on hop rates.

In summary, we introduced  a new  kind of flux balance condition, namely  MB, to obtain steady state   weights  of  nonequilibrium systems  and  demonstrate  its utility in many different kinds  of non-equilibrium dynamics,  including those where the interactions extend beyond two sites. We   believe that 
the  MB technique will be  very helpful in finding steady states  of many other nonequilibrium systems.

\ack The author would like to gratefully acknowledge P. K. Mohanty for his constant encouragement and careful reading of the manuscript. His insightful and constructive comments have helped a lot in improving this work. The author would like to acknowledge the support provided by Saha Institute of Nuclear Physics, where the part of the work has been done. The author also acknowledges the support of Council of Scientific and Industrial Research, India in the form of a Research Fellowship 
(Grant No. 09/921(0335)/2019-EMR-I).

\appendix
\setcounter{section}{1}
\section*{Appendix}
In this Appendix, we study ZRP in two dimensions and obtain a condition on hop rates so that the steady state is factorized. We write the master equation of ZRP in a periodic two dimensional lattice of size $(L \times L)$, sites $(i,j)$ with $i=1,2,3 \cdots L$, $j=1,2,3,\cdots L$ and denote the configurations in terms of the site variables $\{n_{i,j}\}$
\begin{eqnarray}
\fl \frac{d}{dt} P(\{n_{i,j}\})= \sum_{i=1}^{L} \sum_{j=1}^{L} \left[~u_{r}(n_{i-1,j}+1) P(\dots n_{i-1,j}+1,n_{i,j}-1\dots ) \right. \cr \left. \hspace*{1.5cm}+ u_l(n_{i+1,j}+1)P(\dots n_{i,j}-1,n_{i+1,j}+1\dots)
\right. \cr \left. \hspace*{1.5cm}+
u_u(n_{i,j-1}+1)P(\dots n_{i,j-1}+1, \dots n_{i,j}-1\dots)\right. \cr \left. \hspace*{1.5cm}+
u_d(n_{i,j+1}+1)P(\dots n_{i,j}-1\dots n_{i,j+1}+1 \dots)~
\right]\cr
\hspace*{0.5cm} -\sum_{i=1}^{L} \sum_{j=1}^{L}\left[~ u_{r}(n_{i,j})+u_{l}(n_{i,j})+u_{u}(n_{i,j})+u_{d}(n_{i,j}) ~\right] \,\,P(\{n_{i,j}\}) 
\label{eq:master_ZRP} 
\end{eqnarray}
We will like to see if there can be a FSS which neither satisfies DB nor satisfies PWB. In particular, we want to see if there is a FSS which can be obtained using  MB. With a FSS, the steady state master equation for any arbitrary configuration of this two dimensional ZRP model reads as,
\be
\bearr{c}
\fl \sum_{i=1}^{L} \sum_{j=1}^{L} \left[~u_r(n_{i,j})+u_l(n_{i,j})+u_u(n_{i,j})+u_d(n_{i,j})~\right] \,\\ \fl \dots f(n_{i,j-1})\dots f(n_{i-1,j})f(n_{i,j})f(n_{i+1,j})\dots f(n_{i,j+1}) \, \\
\fl -[~\,\sum_{i=1}^{L} \sum_{j=1}^{L} u_r(n_{i-1,j}+1)\dots f(n_{i-1,j}+1)f(n_{i,j}-1)\dots \cr  \fl + \sum_{i=1}^{L} \sum_{j=1}^{L} u_l(n_{i+1,j}+1)\dots f(n_{i,j}-1)f(n_{i+1,j}+1)\dots
\cr \fl +\sum_{i=1}^{L} \sum_{j=1}^{L} u_u(n_{i,j-1}+1) \dots f(n_{i,j-1}+1) \dots f(n_{i,j}-1)\dots
\cr \fl +\sum_{i=1}^{L} \sum_{j=1}^{L} u_d(n_{i,j+1}+1) \dots f(n_{i,j}-1) \dots f(n_{i,j+1}+1) \dots~]=0. \\
 \\\label{eq:master_ZRP1}
\enarr
\ee
One can show following Eq. (\ref{eq:master_ZRP1}) that it is possible to obtain an exact FSS for this model when
\begin{equation}
 u_{r}(n)+u_{l}(n) + u_{u}(n)+u_{d}(n) = u(n) \label{eq:App_u}
\end{equation}
and the steady state is defined as
$f(n) = \prod_{\nu=1}^{n} u(\nu)^{-1}$, if the asymmetric rate functions have the following generic functional form 
\begin{eqnarray}
\fl u_{r}(n) = (u(n)/2)~ [\alpha_{1} -\beta_{1} u(n-1)] ~{\rm ;}~~
u_{l}(n) = (u(n)/2)~ [1-\alpha_{1} + \beta_{1} u(n-1)] \label{eq:fn_url},\\
\fl u_{u}(n) = (u(n)/2)~ [\alpha_{2} -\beta_{2} u(n-1)] ~{\rm ;}~~
u_{d}(n) = (u(n)/2)~ [1-\alpha_{2} + \beta_{2} u(n-1)] \label{eq:fn_uud},
\end{eqnarray}
characterized by  the independent  parameters $0\le \alpha_{1} \le 1$,  $0 \le \beta_{1} \le \alpha_{1}/ u(n)|_{max}$, $0\le \alpha_{2} \le 1$ and  $0 \le \beta_{2} \le \alpha_{2}/ u(n)|_{max}$. The range of $\alpha_{1}$, $\beta_{1}$ and $\alpha_{2}$, $\beta_{2}$ are chosen such that the hop rates $u_{r,l,u,d}(n)$ remain  positive. Note that Eq. (\ref{eq:App_u}) is the most generalized condition on the hop rates to obtain the FSS using MB. The specific choices of hop rates in Eqs. (\ref{eq:2dzrp_cond2}) and (\ref{eq:2dzrp_cond3})  
for the model described in the text also satisfy this generalized condition in Eq. (\ref{eq:App_u}).

\end{document}